\documentclass{emulateapj}  
\usepackage{courier,graphicx,epsfig,natbib,url,twoopt,color}
\usepackage{hyperref} 
\usepackage{pdfcomment}       
\usepackage{acronym}          
\hypersetup{
  colorlinks=true,  
  urlcolor=blue,  
  linkcolor=red  
}
\newdateformat{mydate}{\THEYEAR,\monthname[\THEMONTH], \THEDAy}
\makeatletter
\long\def\frontmatter@title@above{
  \vspace*{-\headsep}\vspace*{\headheight}
   Accepted by {\sc The Astrophysical Journal Letters} 
  \par\vspace*{-\baselineskip}\vspace{12mm}
  }

\makeatother

\def\specchar#1{{\sc{#1}}}    
\def\rmit#1{#1}               
\def\kms{\hbox{km$\;$s$^{-1}$}}
\def\deg{\hbox{$^\circ$}}       
\def\pmj{penumbral microjet}

\def\ie{\rmit{i.e.,}}              
\def\eg{\rmit{e.g.,}}              

\def\Halpha{\mbox{H\hspace{0.1ex}$\alpha$}} 
\def\FeI{\mbox{Fe\,\specchar{i}}} 
\def\CaII{\mbox{Ca\,\specchar{ii}}} 
\def\CaIIH{\mbox{Ca\,\specchar{ii}\,\,H}}
\def\CaIR{\mbox{Ca\,\specchar{ii}\,8542\,\AA}} 
\def\CII{\mbox{C\,\specchar{ii}}} 
\def\MgII{\mbox{Mg\,\specchar{ii}}} 
\def\MgIIk{\mbox{Mg\,\specchar{ii}\,\,k}}

\def\MgIIhk{\mbox{Mg\,\specchar{ii}\,\,h\,\&\,k}}
\def\SiIV{\mbox{Si\,\specchar{iv}}}
\def\OIV{\mbox{O\,\specchar{iv}}} 

\bibpunct{(}{)}{;}{a}{}{,}    
\makeatletter
 \newcommandtwoopt{\citeads}[3][][]{%
   \nonstopmode
   \href{http://adsabs.harvard.edu/abs/#3}%
        {\def\hyper@linkstart##1##2{}%
         \let\hyper@linkend\@empty\citealp[#1][#2]{#3}}
   \biblink{#3}{\href{http://adsabs.harvard.edu/abs/#3}{ADS}}%
   \errorstopmode}            
 \newcommandtwoopt{\citepads}[3][][]{%
   \nonstopmode
   \href{http://adsabs.harvard.edu/abs/#3}%
        {\def\hyper@linkstart##1##2{}%
         \let\hyper@linkend\@empty\citep[#1][#2]{#3}}
   \biblink{#3}{\href{http://adsabs.harvard.edu/abs/#3}{ADS}}%
   \errorstopmode}            
 \newcommandtwoopt{\citetads}[3][][]{%
   \nonstopmode
   \href{http://adsabs.harvard.edu/abs/#3}%
        {\def\hyper@linkstart##1##2{}%
         \let\hyper@linkend\@empty\citet[#1][#2]{#3}}
   \biblink{#3}{\href{http://adsabs.harvard.edu/abs/#3}{ADS}}%
   \errorstopmode}            
 \newcommandtwoopt{\citeyearads}[3][][]{%
   \nonstopmode
   \href{http://adsabs.harvard.edu/abs/#3}%
        {\def\hyper@linkstart##1##2{}%
         \let\hyper@linkend\@empty\citeyear[#1][#2]{#3}}
   \biblink{#3}{\href{http://adsabs.harvard.edu/abs/#3}{ADS}}%
   \errorstopmode}            
\makeatother

\makeatletter
\newcommand{\bibnote}[2]{\@namedef{#1note}{#2}}
\newcommand{\biblink}[2]{\@namedef{#1link}{#2}}
\makeatother

\long\def\rev#1{#1}           

\newcommand{\gregalemail}{g.j.m.vissers@astro.uio.no}

\begin{document}

\title{Evidence for a transition region response to penumbral microjets in
sunspots}
 
\author{G. J. M. Vissers}
\author{L. H. M. Rouppe van der Voort}
\author{M. Carlsson} 
\affil{Institute of Theoretical Astrophysics, University of Oslo, %
  P.O. Box 1029 Blindern, N-0315 Oslo, Norway; 
\gregalemail}

\shorttitle{Transition Region response to Penumbral Microjets}
\shortauthors{Vissers et al.}

\begin{abstract} 
Penumbral microjets are short-lived, fine-structured and bright jets that are
generally observed in chromospheric imaging of the penumbra of sunspots.
Here we investigate their potential transition region signature, by combining
observations with the Swedish 1-m Solar Telescope (SST) in the \CaIIH\ and
\CaIR\ lines with ultraviolet imaging and spectroscopy obtained with the
Interface Region Imaging Spectrograph (IRIS), which includes the
\CII~1334/1335\,\AA, \SiIV~1394/1403\,\AA\ and \MgIIhk~2803/2796\,\AA\ lines.
We find a clear corresponding signal in the IRIS \MgIIk, \CII\ and \SiIV\ slit-jaw
images, typically offset spatially from the \CaII\ signature in the direction
along the jets: from base to top, the penumbral microjets are predominantly
visible in \CaII, \MgIIk\ and \CII/\SiIV, suggesting progressive heating to
transition region temperatures along the jet extent.
Hence, these results support the suggestion from earlier studies
that penumbral microjets may heat to transition region temperatures.
\end{abstract}


\section{Introduction}\label{sec:introduction}
Penumbral microjets (PMJs) are a prime example of the fine-structured dynamics
observed in active regions and in sunspots in particular.
They were first reported by 
\citetads{2007Sci...318.1594K}  
in \CaIIH\ imaging of a sunspot observed at different viewing angles
with the Solar Optical Telescope (SOT; 
\citeads{2008SoPh..249..167T}) 
aboard Hinode
\citepads{2007SoPh..243....3K} 
and were found to occur near bright penumbral grains.
In regular imaging, but especially in time-difference movies, they pop up
unmistakably as small-scale, short-lived jets with average lengths between
1--4\,Mm, widths of a few hundred kilometers, lifetimes of up to a minute and
apparent extension speeds well over 100\,\kms.
PMJs have also been observed in the wings of the \CaIR\ line, with a typically
blue-over-red asymmetric spectral profile peaking between 10--20\,\kms\ blue-ward
of the line core
(\citeads{2013ApJ...779..143R};  
\cite{2014Drews}). 

When observed at different viewing angles, their orientation with respect to the
background penumbra is also revealing: towards the limb microjets appear at
an angle to the photospheric penumbral filaments, while they are well-aligned
closer to disc center.
This suggests that PMJs are (at least initially) more upright than the
photospheric penumbral filaments and indeed inclination angles to
the normal ranging between 20\deg--70\deg\ have been found
\citepads{2007Sci...318.1594K},  
with a tendency for increasing angles from the inner to the outer penumbral boundary
\citepads{2008A&A...488L..33J}. 
This outward trend is indicative of PMJs following the magnetic field
as it expands from the initially near-vertical orientation in the photosphere to
a more horizontal orientation in the chromosphere.

These observations also fit with a scenario in which the PMJs are driven
by reconnection between penumbral magnetic fields.
The darker penumbral filaments are known to host more horizontal fields,
while the brighter penumbral areas and grains harbor more vertical fields
\citepads{2005A&A...436.1087L}.  
As the latter move inwards, towards the sunspot umbra, the shearing reconnection
between inclined penumbral fields may provide the necessary energy to drive the
PMJs.
\citetads{2010A&A...524A..20K}  
and 
\citetads{2010A&A...524A..21J}  
found downflows in the penumbra that in some cases appeared to be related to
PMJs, fitting the reconnection scenario.

The launch of the Interface Region Imaging Spectrograph (IRIS;
\citeads{2014SoPh..289.2733D}) 
has opened a new, high-resolution window on the chromosphere and transition
region and its data are therefore perfectly suited for investigating the
potential transition region response to PMJs (the presence of which was already
speculated by
\citeads{2007Sci...318.1594K}).  
By combining IRIS observations with data from the Swedish 1-m Solar Telescope
(SST;
\citeads{2003SPIE.4853..341S}) 
for PMJ detection, this work aims at investigating the characteristic
ultraviolet signature of the jets and the time-dependence of their visibility in
chromospheric and transition region diagnostics. 

\begin{figure*}
  \centerline{\includegraphics[width=\textwidth]{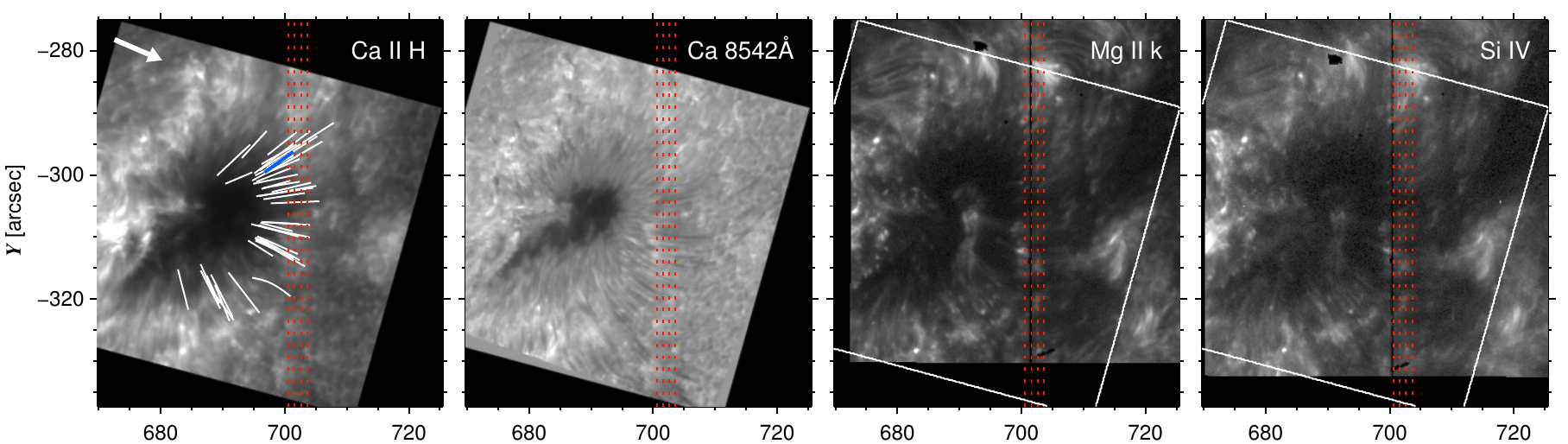}}
  \centerline{\includegraphics[width=\textwidth]{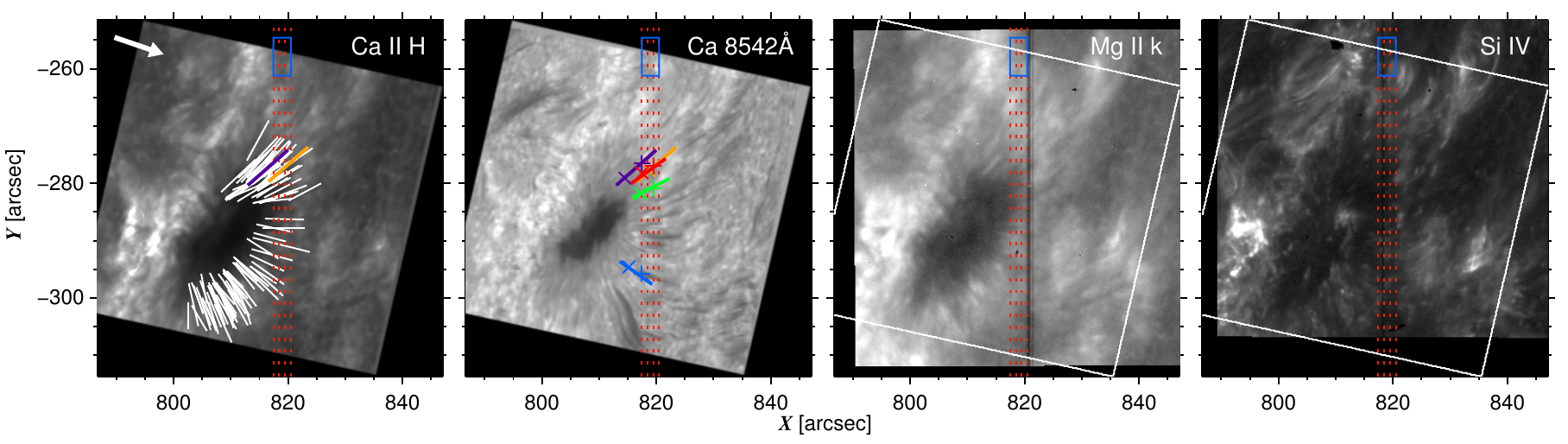}}
  \caption[]{\label{fig:fovs} %
  Field-of-view images for September 5 ({\it top
  row\/}) and September 6 ({\it bottom row\/}).
  {\it Left to right:\/} \CaIIH, \CaIR\ at $-$0.3\,\AA,
  \MgIIk\ and \SiIV\ slit-jaws.
  The vertical dashed red lines specify locations and spacing of the IRIS
  raster positions.
  The arrow in the left-hand panels indicates the direction to the nearest limb.
  The solid white lines in the first column highlight \pmj\ 
  occurrences and the paths along which space-time diagrams have been
  extracted; the colored lines indicate PMJs for which the time evolution is
  shown in Fig.~\ref{fig:timeseries}.
  The colored tracks in the second lower panel specify PMJs for which spectra,
  as sampled at the markers, are shown in Fig.~\ref{fig:spectra}.
  The blue box in the lower panels specifies the averaging region for the mean
  spectrum in that same figure.
  }
\end{figure*}

\section{Observations and Analysis}\label{sec:observations}
\paragraph{Data acquisition and reduction}
For this study we analyzed two data sets of coordinated SST and IRIS
observations obtained on 2014 September 5 and 6, tracking the leading sunspot
in AR~12152.
Figure~\ref{fig:fovs} shows sample field-of-view images for both days in
different SST and IRIS diagnostics.
The CRisp Imaging SpectroPolarimeter (CRISP;
\citeads{2008ApJ...689L..69S}) 
at the SST recorded imaging spectroscopy in the \Halpha\ and \CaIR\ lines, in
addition to full Stokes polarimetry at one wavelength position in the blue wing
of \FeI~6301\,\AA.
The overall repeat cadence of this program is 11.6\,s.
CRISP observed from 07:58--09:52\,UT and from 08:23--10:24\,UT on the
respective days.
In addition, part of the light from the telescope was diverted before entering
CRISP to provide \CaIIH\ core (filter full-width at half-maximum of 1.1\,\AA) and wide-band imaging.
On September 5, the \CaIIH\ data were obtained only during 07:59--09:07\,UT,
while on September 6 the full CRISP sequence was supported by \CaIIH\
imaging.
In this study, only \CaIR\ and \CaIIH\ core data were used for further
analysis.

During these coordinated observations, IRIS ran a 4-step \rev{sparse} raster (OBSID
3820255167) with 2\,second exposures, wavelength binning by 2 in the
far ultraviolet (affecting only the spectra between 1332--1407\,\AA), and
including context slit-jaw imaging in three channels, \CII~1330\,\AA,
\SiIV~1400\,\AA\ and \MgIIk~2796\,\AA.
The overall repeat cadence of both the raster and the slit-jaws is 13\,s.
IRIS co-observed between 08:06--11:06\,UT and 08:05--11:01\,UT on the respective
days.

We used the CRISPRED pipeline 
(see \citetads{2015A&A...573A..40D} 
and references therein) for reduction of the SST/CRISP data. 
The \CaIIH\ data were processed using a similar set of procedures, including
dark and flat field corrections, as well minimization of high-order seeing
effects using Multi-Object Multi-Frame Blind Deconvolution
(MOMFBD; \citeads{2005SoPh..228..191V}) 
and removal of rubber-sheet distortions through destretching following
\citetads{1994ApJ...430..413S}. 
The resulting \CaIIH\ data have the same 11.6\,s cadence as the CRISP data.
Co-alignment was achieved using the procedures as described in 
\citetads{2015arXiv150700435V}  
\rev{and using SST \CaIR/\CaIIH\ (for September 5 and 6, respectively) and IRIS
\MgIIk~2796\,\AA\ as reference}.  
The CRisp SPectral EXplorer (CRISPEX;
\citeads{2012ApJ...750...22V}) 
was extensively used during alignment and for subsequent data analysis.

\paragraph{Event selection}
Penumbral microjets were initially detected based on morphology and lifetime
dynamics through visual inspection of \CaIIH\ time-difference sequences
constructed with a temporal offset of 3 time-steps (about 35\,s).
This follows the methodology that has been employed in several PMJ studies
before (\eg\
\citeads{2007Sci...318.1594K};  
\citeads{2008A&A...488L..33J}). 
These identifications were subsequently checked against the regular \CaIIH\
imaging to confirm their PMJ nature.
\rev{We restricted our event sample for final analysis to occurrences when IRIS
was not affected by increased cosmic ray hits (resulting from the orbital pass
through the South Atlantic Anomaly), yielding a grand total of 180 events (along
147 paths, \ie\ counting recurring jets separately), of which 52 were on
September 5 and 128 on September 6.}
This selection of PMJs is highlighted by solid lines in the first column
of Fig.~\ref{fig:fovs}. 

\begin{figure*}
  \centerline{\includegraphics[width=\textwidth]{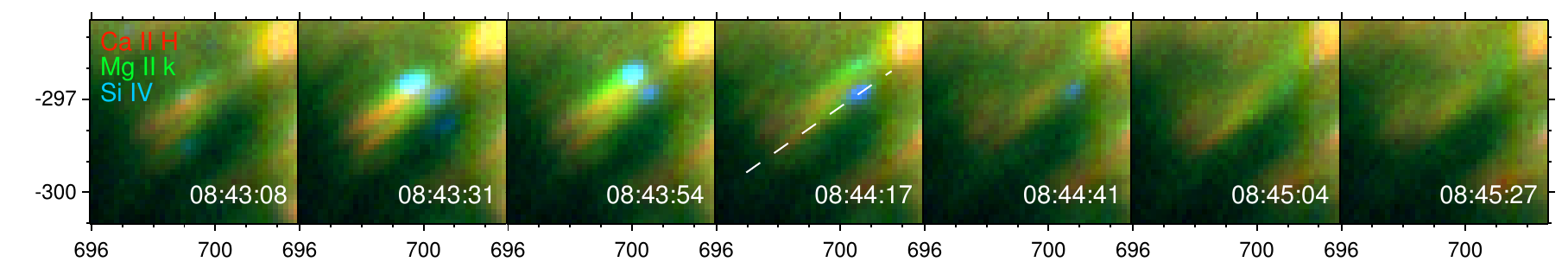}}
  \centerline{\includegraphics[width=\textwidth]{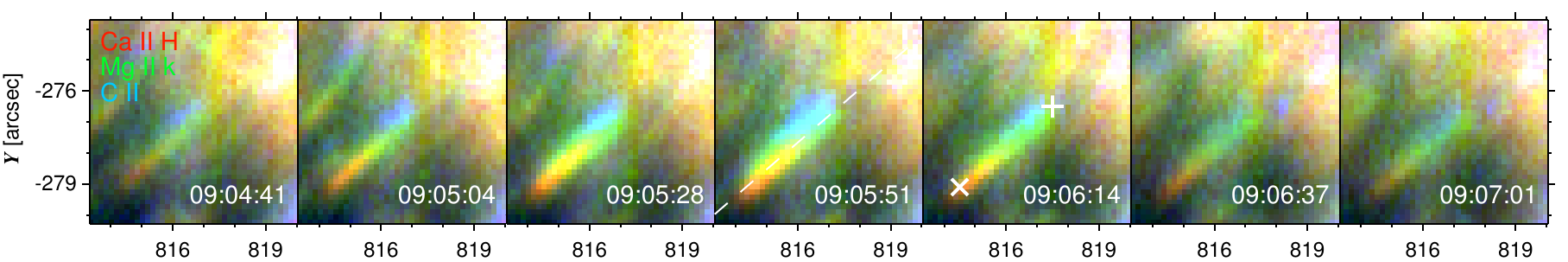}}
  \centerline{\includegraphics[width=\textwidth]{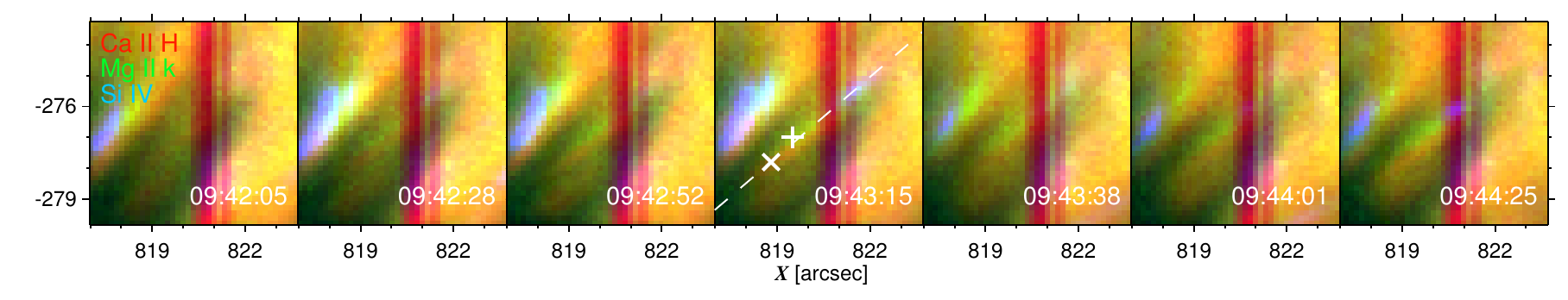}}
  \caption[]{\label{fig:timeseries} %
  Time evolution of three PMJ examples.
  From top to bottom these correspond to the blue, violet and orange tracks 
  highlighted in the left-hand column of Fig.~\ref{fig:fovs}.
  The diagnostic color coding is indicated in the top left of each
  panel, while the time in UT is specified at the bottom.
  The dashed white lines specify the paths along which space-time diagrams
  (b), (f) and (h) (shown in Fig.~\ref{fig:timeslices}) have been extracted, while
  the symbols correspond to sampling locations \rev{and timing} of the violet
  and orange \rev{\CaIR\ ({\textit{cross marker\/}}) and IRIS ({\textit{plus
  marker\/}})} spectra shown in Fig.~\ref{fig:spectra}.
  \rev{Note that the bright event to the left in the bottom example resulted
  from multiple fine-structure loops brightening (distinguishable only at SST
  resolution) and exhibited ``IRIS bomb''-type \citepads{2014Sci...346C.315P} UV
  profiles, in contrast to all PMJ spectra analyzed; it was therefore discarded
  as microjet in our analysis.}
  }
\end{figure*}

\section{Results}\label{sec:results}
\paragraph{Morphology and time evolution}
Figure~\ref{fig:timeseries} shows the time evolution of three example PMJs that
are highlighted by colored tracks in the first column of Fig.~\ref{fig:fovs}.
Their visibility in different diagnostics is indicated through color coding:
\CaIIH\ signal is red, \MgIIk\ is green and \SiIV\ or \CII\ is blue.
For events that are visible in multiple channels the colors add; 
a feature visible in \CaIIH\ and \MgIIk\ turns yellow, in \MgIIk\
and \SiIV\ more cyan, while something visible in all three tends to
white.
Using nearest-neighbor selection, the time differences between the diagnostics
are small, but non-zero nevertheless: about 3.3\,s on average and 6.7\,s at
most between the IRIS slit-jaws and SST, and 6.4\,s and 9.8\,s between the
\MgIIk--\SiIV\ and \MgIIk--\CII\ pairs, respectively.

Although their lifetimes, sizes and brightness differ, these examples display a
generally observed rainbow-colored signature. 
The signal in \CaIIH\ (or \CaIR, not shown here) is predominantly found towards
the base of the PMJ, \MgIIk\ is observed throughout but mostly in the middle,
while \SiIV\ and \CII\ are generally only observed towards the PMJ tops.
While the \CaII\ and \MgIIk\ signal are elongated and jet-like in shape, the
\SiIV\ and \CII\ signature is more often roundish.
Given the viewing angle, this is unlikely an effect of foreshortening, as the
jets are viewed from the side.
The \CII\ PMJ signal appears somewhat more elongated than \SiIV\ and is
sometimes also offset spatially;
for some events this may be an intrinsic effect, as the offset is in the
direction perpendicular to the extension of the jet, but for most cases it seems
likely a result of the non-negligible 3.5\,s time delay between the \CII\ and
\SiIV\ exposures.

\begin{figure*}
  \centerline{\includegraphics[width=\textwidth]{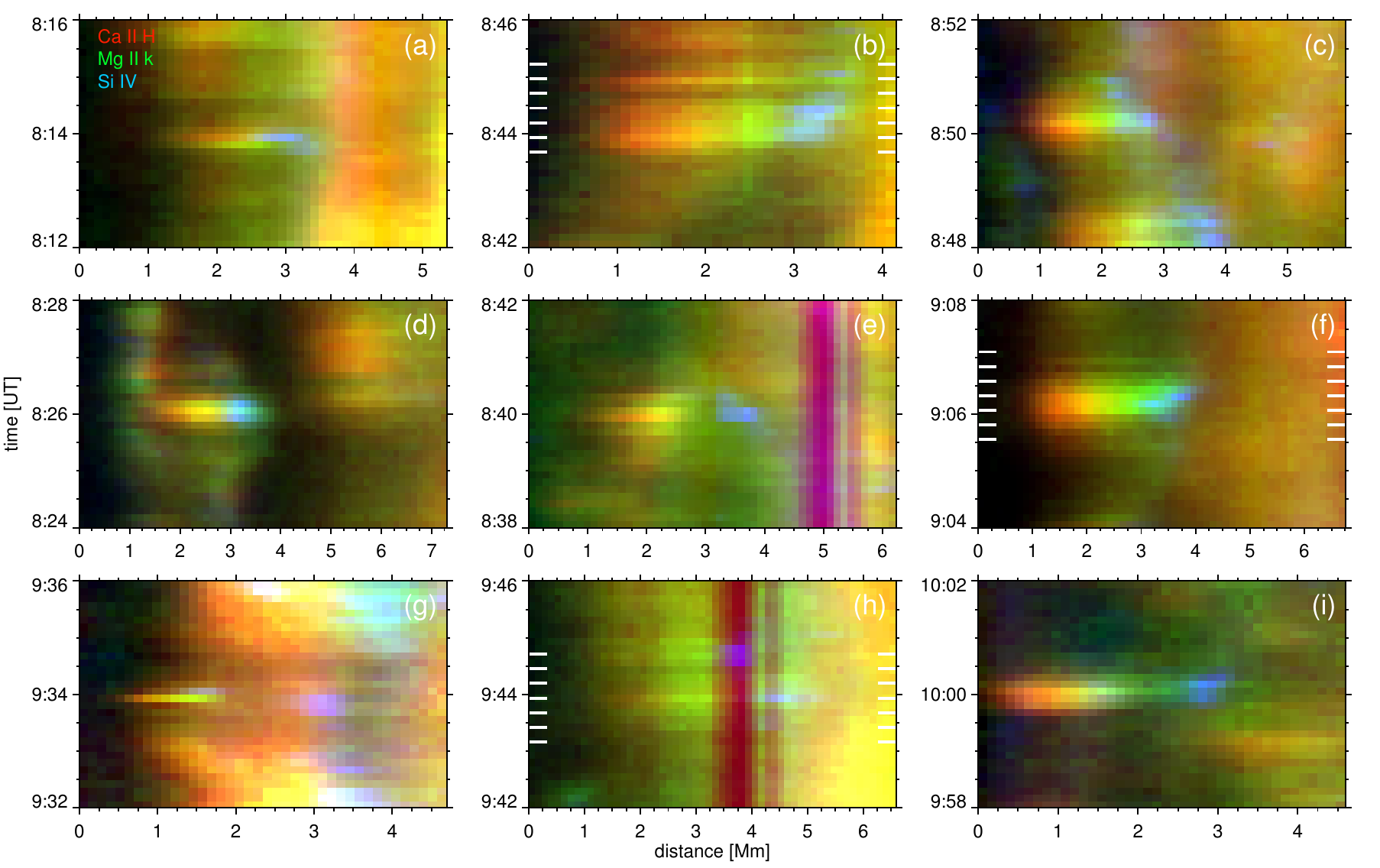}}
  \caption[]{\label{fig:timeslices} %
  Space-time diagrams for nine representative PMJs on September 5, panels
  (a)--(c), and September 6, panels (d)--(i).
  Color coding convention as for Fig.~\ref{fig:timeseries}.
  The thick white inner tickmarks in panels (b), (f) and (h) correspond to the
  times for which Fig.~\ref{fig:timeseries} shows cutout intensity images.
  }
\end{figure*}

Space-time diagrams were extracted along all paths specified in
Fig.~\ref{fig:fovs}, in order to further analyze the PMJ time evolution.
Figure~\ref{fig:timeslices} presents a representative subset of those.
The rainbow-colored pattern is clearly visible in all examples, but the details
differ from case to case.
For instance, in panel (a) the \MgIIk\ signal appears to be strongly present
prior to the visibility in \CaIIH\ and \SiIV, while in all other panels the
\MgIIk\ and \CaIIH\ signal is roughly co-temporal.
Indeed, in most panels the PMJ appears at the same time or within one time step
in all three diagnostics.
Panels (b) and (c) suggest jet recurrence on the order of 1--2\,min along the
same path and the latter panel also evidences the radial inward migration of PMJs
that was already shown by 
\citetads{2007Sci...318.1594K}.  
The \SiIV\ signal in panels (c) and (e) appears to retract during the PMJ
lifetime (with apparent speeds of about 2--4\,\kms), while the opposite
apparent motion (at about 10\,\kms) is clearly visible in for instance panel
(f).
Some PMJs are only visible for 10--20\,s (\eg\ panels (a), (g) and (h)), while
others remain visible for up to a minute (\eg\ panels (e) and (f)).

\begin{figure}
  \centerline{\includegraphics[width=\columnwidth]{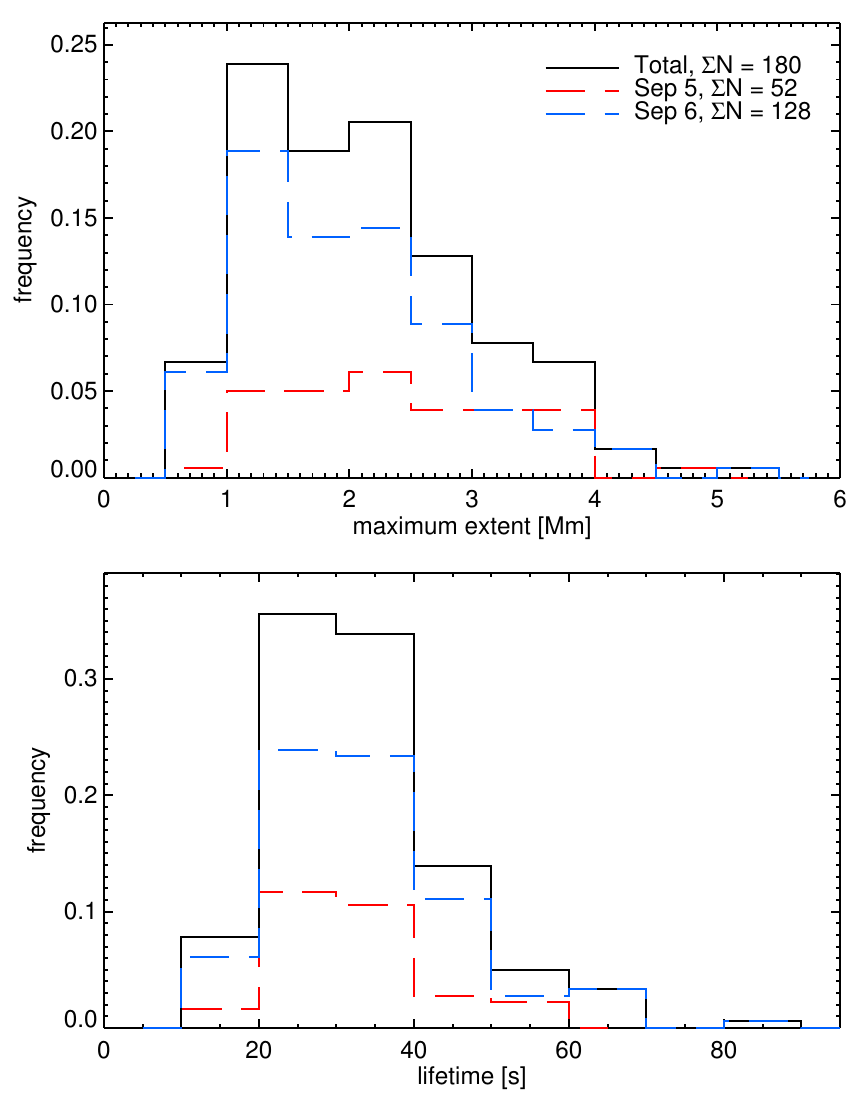}}
  \caption[]{\label{fig:stats} %
  Distribution of maximum PMJ extent ({\it \rev{top}\/}) and visibility lifetime
  ({\it \rev{bottom}\/}) as determined from the space-time diagrams of all PMJs shown in
  Fig.~\ref{fig:fovs}.
  Different line styles indicate different samples, as indicated in the top
  right of the \rev{top} panel.
  }
\end{figure}

Figure~\ref{fig:stats} visualizes these properties statistically, showing the
frequency distribution of maximum extent and lifetimes.
Both represent the combined visibility in \CaIIH, \MgIIk\ and \SiIV, which
especially in terms of spatial extent is appreciably larger than for \CaIIH\
alone (as evident from Fig.~\ref{fig:timeslices}).
We find typical lengths ranging between about 1--5\,Mm for PMJs from both data
sets, with an overall average of 2.1\,Mm, however nearly 57\% of the PMJs are
shorter than that.
The lifetime distribution is equally skewed to lower values, ranging between
10--90\,s and with an average of slightly over half a minute.  
Close to 78\% of the PMJs are visible for 35\,s or less.

\paragraph{Spectral signature}
Although many of the traced paths in Fig.~\ref{fig:fovs} are partially covered
by the IRIS raster, the number of events for which spectra are available is
limited (note that the traced paths indicate the track along which space-time
diagrams have been obtained, not the actual extent of the PMJs).
Not only do not all PMJs along those tracks reach the left-most raster position,
but of those that do, the \MgII, \SiIV\ or \CII\ slit-jaw signal is not always
strong, resulting in an only weakly enhanced spectral signal.
\rev{For September 5, 10 events show a response in either \CII/\SiIV\ or
\MgIIhk, and only one is enhanced in all lines; for September 6, 12 events show
response in either lines and 7 are enhanced in all of them.}
Nonetheless, the IRIS spectra show that the visibility of PMJs in the \MgIIk,
\SiIV\ and \CII\ slit-jaw images is a result of an enhancement of the spectral
lines dominating their passbands. 

\begin{figure*}
  \centerline{\includegraphics[width=\textwidth]{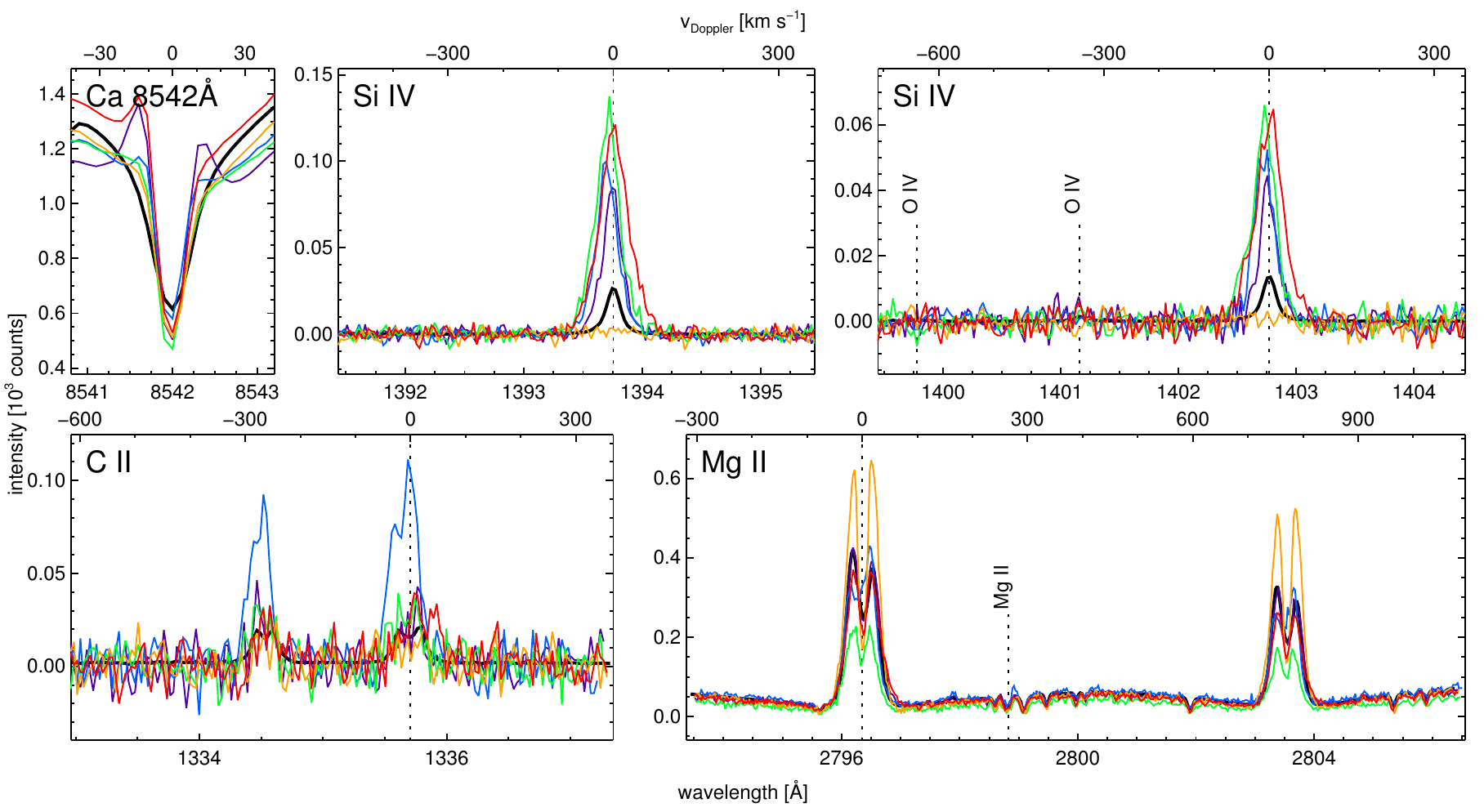}}
  \caption[]{\label{fig:spectra} %
  Sample spectra for five PMJs highlighted with the same color-coding in
  the second lower panel of Fig.~\ref{fig:fovs}.
  For each, the IRIS spectra have been obtained at the same location ({\it plus
  markers\/}, +), while the \CaIR\ stem from the base of the PMJ ({\it cross
  markers\/}, $\times$).
  The thick solid black line is the average over the blue box in
  Fig.~\ref{fig:fovs} (for the IRIS spectra) or the full SST field-of-view (for
  the \CaIR\ line).
  }
\end{figure*}

Figure~\ref{fig:spectra} shows sample spectra for five PMJs from September 6
with the strongest IRIS response and which are highlighted with the same
color-coding in the second lower panel of Fig.~\ref{fig:fovs}.
In addition, the violet and orange curves correspond respectively to the PMJs
shown in the middle and lower panels of Fig.~\ref{fig:timeseries}, and the (f)
and (h) space-time diagrams in Fig.~\ref{fig:timeslices}.

For all PMJs, the \CaIR\ line shows the typical blue-over-red wing asymmetry, 
peaking out at 10--15\,\kms\ blue-ward of the nominal line center. 
The red wing is generally smooth with intensity close to the field-of-view
average, but the blue and violet \rev{samplings} also show a red wing
enhancement at similar Doppler shift, albeit still lower in intensity than the
blue wing peak.

Excluding the orange sampling (which was chosen to show the \MgIIhk\ response),
the \SiIV\ lines appear the most enhanced of the IRIS diagnostics.
The peak ratio of the \SiIV\,1394\,\AA\ to 1403\,\AA\ lines varies between
1.8--2.0, close enough to their transition probability of 2 to suggest 
optically thin formation for all samplings shown.
The behavior of both \SiIV\ lines is also similar: both 
profiles rise by a factor of 3--5 with respect to the quiescent profiles, while
broadening appreciably (up to almost a factor 2 for the broadest red profiles).
With the exception of the blue and green profiles, which show blue shifts of
about 10\,\kms\ with respect to the average, the \SiIV\ lines are generally
symmetric and show close to zero Doppler shift.
There is no sign of \OIV~1399.78\,\AA\ or 1401.16\,\AA\ signal 
(labeled in the top right panel) for any of the samplings.

By comparison, the \CII\ lines appear rather docile.
Only the blue sampling shows considerable enhancement over the quiescent
profile, while all other samplings are dominated by noise.
The blue profiles show a hint of red-asymmetry, which is not reflected by the
\SiIV\ lines, but which is visible in both the \MgIIhk\ lines as well as the
\MgII\ triplet lines at 2798.82\,\AA\ (indicated by the labeled vertical dashed
line in the lower right panel).
Comparison with the images and spectra in \MgIIk, \CaIR\ and \Halpha\ suggests
this is due to absorption by an overlying fibril that has been blue-shifted by
the inverse Evershed effect.

As Figs.~\ref{fig:timeseries} and \ref{fig:timeslices} already suggest, the
\SiIV\ or \CII\ signal is generally not co-located with the \MgIIk\ signal.
Hence, it is not surprising that the response in \MgIIhk\ is generally weak (if
at all present) in most samplings shown in Fig.~\ref{fig:spectra}.
Indeed, the response in the \SiIV\ and \MgIIk\ lines appears to be
anti-correlated at these sampling locations: the stronger the \SiIV\ (increasing
for violet, blue, red to green), the weaker the \MgIIhk.
Conversely, the orange sampling, which has been selected specifically to show
the PMJ \MgIIhk\ response, shows nothing but noise at the rest-wavelengths of
both \SiIV\ lines.
Compared to the quiescent profile, both \MgII\ lines increase in intensity only in
the k$_{2V/R}$ and h$_{2V/R}$ peaks; the k$_{3}$/h$_{3}$ cores
appear in fact darker at this sampling location than the average profile.
Although both lines get broadened, the effect is not as strong as for
\SiIV.

\section{Discussion and Conclusions}\label{sec:discussion}
We have presented a study of the transition region signature of
penumbral microjets in coordinated observations of a sunspot by the SST and IRIS
on September 5--6, 2014.
Our key finding is the progressive visibility of PMJs in chromospheric and
transition region diagnostics along their extent: from base to top the jets are
mainly visible in \CaIIH, \MgIIk\ and \SiIV/\CII.
These results support the suggestion by
\citetads{2007Sci...318.1594K}  
that PMJs may have a transition region component.

The time delay between visibility in these diagnostics is close to negligible,
differing at most by about 10\,s.
Given the characteristic formation temperatures of these lines (10$^{4.0}$\,K
for \MgIIk, 10$^{4.3}$\,K for \CII\ and 10$^{4.8}$\,K for \SiIV), as well as the
spatial offsets along the jet direction (typically at least 500--1000\,km), there
is likely to be an intrinsic time delay in the diagnostic visibility, however
the current cadence does not allow disentangling these conclusively.

The PMJ-related signal in \MgIIk\ and \SiIV/\CII\ generally follows the
extension set by the \CaIIH\ jet; of all 147 cases considered, only two
showed some curvature of their path.
Such behavior was already reported by 
\citetads{2008A&A...488L..33J}, 
who found a small angle between the initial and final orientation of a PMJ in
consecutive frames.
Even though many PMJs appear slender throughout, we find
that some jets may broaden up to about 500\,km
towards the top.
The maximum extents (\ie\ combining \CaIIH, \MgIIk\ and \SiIV\ signal) that we find
are of the order of the values reported for \CaIIH\ alone by 
\citetads{2007Sci...318.1594K},  
while the \CaIIH\ lengths in our samples appear to be much shorter (about
0.5--1\,Mm, comparable to the values found for \CaIR\ PMJs by 
\citet{2014Drews}). 

Recently, 
\citetads{2014ApJ...790L..29T} 
reported on IRIS observations of bright dots in sunspots, the majority of which
were observed in the penumbrae. 
The dot-like brightenings in their \SiIV\ slit-jaw images appear to
have properties similar to the PMJs in our data:
\rev{they have similar small, round shapes, lifetimes on the order of a minute and
those that show proper motion, have comparable speeds of 10--40\,\kms\ (our PMJs
fall on the lower end of that range).
The IRIS response is also comparable in the enhancement of the
\SiIV~1402.77\,\AA\ and \CII~1334.53\,\AA\ lines, the weak
\MgIIk~2796.35\,\AA\ response, and the absence of significant continuum
enhancement and \OIV~1401.16\,\AA\ signature.}
Tian and co-workers suggested that some of these dots could be related to PMJs, given
the weak but present \MgIIk\ signal, but that many were likely related to
energy releases at loop footpoints (in particular those with counterpart signal
in the coronal channels of 
the Solar Dynamics Observatory's Atmospheric Imaging
Assembly 
\citepads{2012SoPh..275...17L}). 
Considering our findings, we speculate that indeed a large number of their
moving bright dots may have been PMJ tops.

The \CaIR\ spectra from the base of the PMJs generally show a blue-over-red
asymmetry peaking at some 15\,\kms, in accordance with literature values
(\citeads{2013ApJ...779..143R};  
\cite{2014Drews}). 
The interpretation of such profile is not straightforward, 
as was already pointed out by \citetads{2013ApJ...779..143R}.  
The peak may represent an actual Doppler component at that velocity, but 
could also merely be a result of the overlying canopy, as for instance for
Ellerman bombs
(\citeads{2011ApJ...736...71W}; 
\citeads{2013ApJ...774...32V}). 
In at least one example (\ie\ the blue profiles), the latter appears to be the
case.
On the other hand, a few examples of a similar, co-temporal blue-shift velocity
in both \SiIV\ lines and the \CaIR\ line are found, suggesting a persistent
Doppler shift throughout the PMJ.
Whether these are representative, however, is difficult to say without a larger
statistical sample \rev{of IRIS spectra}.
Although the \CaIR\ PMJ profiles are comparable to those of Ellerman bombs, 
the enhancements in the \SiIV, \CII, and \MgIIhk\ lines are no stronger
than the weakest IRIS Ellerman bomb profiles reported in
\citetads{2015arXiv150700435V};  
more often than not they are much weaker than those.

In conclusion, penumbral microjets have a clear corresponding signature in the IRIS
upper chromosphere and transition region diagnostics.
The progressive visibility in \MgIIk\ and \SiIV/\CII\ following the initial jet
in \CaIIH\ and \CaIR, combined with the enhanced and broadened \SiIV\ lines, suggest
heating of plasma to transition region temperatures as it extends upwards (while
also slightly expanding, likely following the expanding canopy fields).
A natural next step would be to expand this study statistically, preferably
by sampling a larger part of the penumbra with the IRIS raster than was done in
these data sets, as well as doing so at a higher cadence to better disentangle
their time evolution.

\acknowledgments %

Our research has been funded by the Norwegian Research
Council and by the ERC under the European Union's Seventh
Framework Programme (FP7/2007-2013)\,/\,ERC grant agreement
nr.~291058.
IRIS is a NASA small explorer mission developed and operated by LMSAL
with mission operations executed at NASA Ames Research Center
and major contributions to downlink communications funded by the
Norwegian Space Center through an ESA PRODEX contract.
The SST is operated on the island of La Palma by the Institute for
Solar Physics of Stockholm University in the Spanish Observatorio del
Roque de los Muchachos of the Instituto de Astrof{\'\i}sica de
Canarias.

\mbox{} 


\end{document}